\documentclass[8pt]{article}
\usepackage{graphicx, array, blindtext}
\usepackage{amsmath}

\begin{document}			

\title{\Large{Interaction between Generalized Varying Chaplygin gas and Tachyonic Fluid}}

\author{\small{Martiros Khurshudyan}\\
\texttt{\small{CNR NANO Research Center S3, Via Campi 213a, 41125 Modena MO, Italy}}\\\texttt{\small{and}}\\
\texttt{\small{Dipartimento di Scienze Fisiche, Informatiche e Matematiche,}}\\
\texttt{\small{ Universita degli Studi di Modena e Reggio Emilia, Modena, Italy}}\\ \\
\texttt{\small{email:martiros.khurshudyan@nano.cnr.it}}}
\date{\small{\today}}
\maketitle
\vspace{10 mm}
\begin{abstract}
We consider a mixture of varying Generalized Chaplyagin gas \cite{ZhangX} interacting with a Tachyonic fluid in framework of GR. We suppose, that a Tachyonic fluid and Generalized Chaplyagin gas was separated somehow from Darkness of the Universe. Remaining darkness will be described by $\omega_{X}$ and we will have not any other information. We also suppose, that varying Generalized Chaplyagin gas under consideration was appeared due to interaction between remaining Darkness and Generalized Chaplyagin gas. At the same time we assume, that our Tachonic fluid does not fell remaining Darkness. In this letter we investigate graphically the potential $V(\phi)$ of the Tachyonic fluid and $\omega_{\small{tot}}$ of the mixture. Interaction between components taken to be sign-changeable, occurring thanks to imposing of deceleration parameter into expression of interaction term \cite{Hao}, \cite{Hao2}.
\end{abstract}
\newpage
\section*{\large{Introduction}}
Scalar fields thought to be one of the ways of describing Dark Energy. Dark Energy was born in GR  to explain experimental data claiming, that the expansion of the Universe is accelerated.
Dark Energy with Dark Matter are exotic and mysterious components of the Universe, capturing a lot of attention to be explained still. Some of the thoughts and attempts are to make link between Dark Energy and Dark Matter with extra-dimensions, because both Dark Components and extra-dimensions, are invisible today. This as well as other ideas should be checked and verified by experiments and as now we have not enough information we make different manipulations concerning to the topic. Generally, it is accepted that Dark Energy should be described by negative pressure and positive energy density, therefore implying negative EoS parameter $\omega<0$. The simplest model is a cosmological constant $\omega_{\Lambda}=-1$ introduced by Einstein, but it suffers with the problem known as cosmological coincidence problem asking why are we living in an epoch in which the densities of dark energy and matter are comparable? One of the ways to solve the cosmological coincidence problem is to consider the interaction between the components on phenomenological level. Generally, interaction could be considered as a function of energy densities and their derivatives: $Q(\rho_{i},\dot{\rho}_{i},\ldots)$. To make everything working with units, we only should consider the fact that unit of interaction term $[Q]=\frac{[energy ~ density]}{time}$ and assume that unit $time^{-1}$ could be contributed, for instance, from Hubble parameter, which gives the forms of the interactions considered in literature: $Q=3Hb\rho_{m}$, $Q=3Hb\rho_{\small{de}}$, $Q=3Hb\rho_{\small{tot}}$, where $b>0$ is a coupling constant, to time dependent forms $Q=\gamma\dot{\rho}_{\small{m}}$, $Q=\gamma\dot{\rho}_{\small{de}}$, $Q=\gamma\dot{\rho}_{\small{tot}}$, an interaction of the general form $Q=3Hb\gamma\rho_{i}+\gamma\dot{\rho_{i}}$, where $i=\{m,de,tot\}$. Interaction between components arose as a result of splitting of the energy conservation, which mathematically can be described as follows:
\begin{equation}
\dot{\rho}_{DE}+3H(\rho_{DE}+P_{DE})=Q
\end{equation}
and
\begin{equation}
\dot{\rho}_{\small{DM}}+3H(\rho_{\small{DM}}+P_{\small{DM}})=-Q,
\end{equation}
which could be understood as: there is not energy conservation for the components separately, but due to interaction between all components, energy of whole mixture conserves. This approach is correct for this moment only.
Among the models connected to the Dark Energy for the present article we will consider a scalar field given by a relativistic Lagrangian and called Tachyonic field \cite{Sen}.
\begin{equation}\label{eq:TF}
L=-V(\phi)\sqrt{1-\partial_{\mu}\phi\partial^{\mu}\phi}.
\end{equation}
Energy density  and pressure of this model in case of FRW metric reads as
\begin{equation}\label{eq:TFdensity}
\rho_{\small{TF}}=\frac{V(\phi)}{\sqrt{1-\dot{\phi}^{2}}}
\end{equation}
and
\begin{equation}\label{eq:TFpressure}
P_{\small{TF}}=-V(\phi)\sqrt{1-\dot{\phi}^{2}}.
\end{equation}
The other considered component is Varying Generalized Chaplygin Gas with EoS
\begin{equation}\label{eq:ChGas}
P_{ncg}=-\frac{A(a)}{\rho^{\alpha}_{ncg}}.
\end{equation}
Where $A$ is a function of scale factor \cite{ZhangX}:
\begin{equation}\label{eq:A}
A(a)=-\omega_{X} A_{0}a^{-3(1+\omega_{X})(1+\alpha)}.
\end{equation}
We assume that somehow, concerning to an unknown physics (even could be very well known) it was possible to separate components of the Darkness of the Universe and was found that we have $TF+GCg+Remaining Darkness$. Only information which we know about Remaining~Darkness (RD) is that its EoS $\omega_{X}<0$. We assume that interaction between Generalized Chaplygin Gas \cite{Chgas} and Remaining Darkness gives to born a Varying Generalized Chaplygin Gas \cite{Chimento} with (\ref{eq:A}) and that there is not any interaction between TF and RD: TF does not feel RD. This assumption allow us still to think that our Universe filled by mixture of a TF and a varying GC gas. Then we will assume that EoS parameter of RD $\omega_{X}$ is a function of time: $\omega(t)=\omega_{0}+\omega_{1}(t\dot{H}/H)$, which has an explicit time dependence that disappears with the $t\dot{H}=H$ condition \cite{Usmani}. Which in its turn gives us a modification of (\ref{eq:A})
\begin{equation}\label{eq:modA}
A(a)=-\omega(t) A_{0}a^{-3(1+\omega(t))(1+\alpha)}.
\end{equation}
The mixture of our consideration will be described by $\rho_{\small{tot}}$ and $P_{\small{tot}}$ given by
\begin{equation}\label{eq:mixture energy}
\rho_{ \small{tot}} = \rho_{ \small{TF}}+\rho_{ \small{ncg}}
\end{equation}
and
\begin{equation}\label{eq:mixture presure}
P_{ \small{tot}} = P_{ \small{TF}}+P_{ \small{ncg}}.
\end{equation}
and interaction between components will be considered $Q=q(3Hb\rho+\gamma\dot{\rho})$, where b and $\gamma$ both dimensionless constants, the energy density $\rho$ could be $\rho_{m}$, $\rho_{\small{de}}$, $\rho_{tot}$ \cite{Hao}, \cite{Hao2}. $q$ is the deceleration parameter
\begin{equation}\label{eq:decparameter}
q=-\frac{1}{H^{2}} \frac{\ddot{a}}{a}.
\end{equation}
$H$ is the Hubble parameter and $a(t)$ is a scale factor.\\ \\
Three cases describing accelerating Universe will be considered here
\begin{enumerate}
   \item Emergent scenario $a(t)=a_{0}(B+e^{Kt})^{m}$, $a_{0}>0, K>0, B>0, m>1$ \cite{Mukherjee}
   \item Intermediate scenarion $a(t)=\exp[\lambda t^{\beta}]$, $\lambda>0$ and $0<\beta<1$ \cite{Barrow1}, \cite{Barrow2}
   \item Logamediate scenario $a(t)=\exp[\mu(\ln(t))^{\alpha}]$, $\mu\lambda>0$, $\alpha>1$ \cite{Barrow1}
\end{enumerate}
Paper organized as follow: Introduction, then we will give field equations and a strategy how to solve and find needed parameters to investigate. In last three sections numerical analysis of $V(\phi)$ and $\omega_{\small{tot}}$ of the Tachyonic Matter will be presented. In section Discussion obtained results will be summarised.
\section*{Field Equations and Mathematics}
Field equations that governs our model of consideration are
\begin{equation}\label{eq:Einstein eq}
R^{\mu\nu}-\frac{1}{2}g^{\mu\nu}R^{\alpha}_{\alpha}=T^{\mu\nu}
\end{equation}
and with FRW metric (the metric of a spatially flat homogeneous and isotropic universe)
\begin{equation}\label{eq:FRW metric}
ds^{2}=dt^{2}-a(t)^{2}\left(dr^{2}+r^{2}d\theta^{2}+r^{2}\sin^{2}\theta d\phi^{2}\right).
\end{equation}
they reduce to
\begin{equation}\label{eq: Fridmman vlambda}
H^{2}=\frac{\dot{a}^{2}}{a^{2}}=\frac{\rho_{\small{tot}}}{3}
\end{equation}
\begin{equation}\label{eq:Freidmann2}
- \frac{\ddot{a}}{a}=\frac{1}{6}(\rho_{\small{tot}}+P_{\small{tot}}).
\end{equation}
Energy conservation condition reads as
\begin{equation}\label{eq:Bianchi eq}
\dot{\rho}_{\small{tot}}+3\frac{\dot{a}}{a}(\rho_{\small{tot}}+P_{\small{tot}})=0.
\end{equation}
To introduce an interaction between Dark Energy and Matter, as already was mentioned in Introduction section, (\ref{eq:Bianchi eq}) splits into two following equations
\begin{equation}\label{eq:inteqm}
\dot{\rho}_{cg}+3H(\rho_{ncg}+P_{ncg})=Q
\end{equation}
and
\begin{equation}\label{eq:inteqG}
\dot{\rho}_{\small{TF}}+3H(\rho_{\small{TF}}+P_{\small{TF}})=-Q.
\end{equation}
In our case problem solving strategy will start with finding a solution of (\ref{eq:inteqm}).
Then the energy density and pressure of a Tachyonic Fluid will be
\begin{equation}\label{eq:tfendensity}
\rho_{\small{TF}}=3H^{2}-\rho_{\small{ncg}}.
\end{equation}
\begin{equation}\label{eq:TFpressure}
P_{\small{TF}}=\frac{-Q-\dot{\rho}_{\small{TF}}}{3H}-\rho_{\small{TF}}.
\end{equation}
Taking into account that EoS parameter of Tachyonic Matter and filed $\phi$ related to each other by the following expression: $\omega_{\small{TF}}=\frac{P_{\small{TF}}}{\rho_{\small{TF}}}=-1+\dot{\phi}^{2}$, for the field we should have
\begin{equation}\label{eq:filed}
\phi=\int{\sqrt{1+\omega_{\small{TF}}}~dt}.
\end{equation}
Potential can be recovered by
\begin{equation}\label{eq:TFpot}
V(\phi)=\sqrt{-\rho_{\small{TF}}P_{\small{TF}}}.
\end{equation}
$"-"$ under the square root, will not worry us, because $P_{\small{TF}}<0$. The other parameter of our interest is $\omega_{\small{tot}}$ defined as
\begin{equation}\label{eq:omegatot}
\omega_{\small{tot}}=\frac{P_{\small{TF}}+P_{ncg}}{\rho_{\small{TF}}+\rho_{ncg}}.
\end{equation}
\section*{\large{Emergent scenario}}
\begin{figure}
\centering
\includegraphics[width=50 mm]{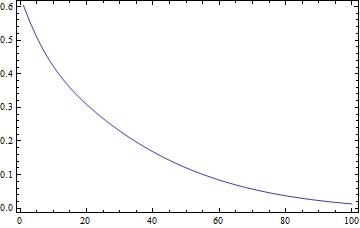}
\caption{%
The variation of $V$ against t, Interaction: $Q=q(3Hb\rho_{ncg}+\gamma\dot{\rho}_{ncg})$,
Parameters: $\omega_{0}=-0.8$, $\omega_{1}=-0.03$, $m=2$, $K=0.03$, $A_{0}=1$, $a_{0}=1$, $B=0.5$, $\gamma=0.1$, $b=0.3$, $\alpha=0.5$} 
\label{fig:potemergent}
\end{figure}

\begin{figure}
\centering
\includegraphics[width=50 mm]{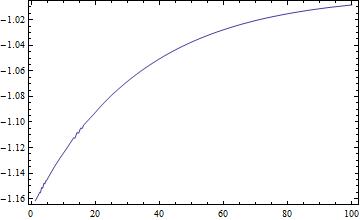}
\caption{%
The variation of $\omega_{\small{tot}}$ against t, Interaction: $Q=q(3Hb\rho_{ncg}+\gamma\dot{\rho}_{ncg})$,
Parameters: $\omega_{0}=-0.8$, $\omega_{1}=-0.03$, $m=2$, $K=0.03$, $A_{0}=1$, $a_{0}=1$, $B=0.5$, $\gamma=0.1$, $b=0.3$, $\alpha=0.5$} 
\label{fig:omegaemergent}
\end{figure}

Performing numerical analysis of the problem in case of Emergent scenario with a scale factor $a=a_{0}(B+\exp[Kt])^{m}$ and Hubble parameter
\begin{equation}\label{eq:HubblEmergent}
H=\frac{Kme^{Kt}}{B+e^{Kt}}.
\end{equation}
we observe that in case with a set of parameters, allowing us to obtain $V\rightarrow0$
over time (Fig. \ref{fig:potemergent}), $\omega_{\small{tot}}<-1,$ which shows phantom-like behavior (Fig. \ref{fig:omegaemergent}).
\begin{figure}
\centering
\includegraphics[width=50 mm]{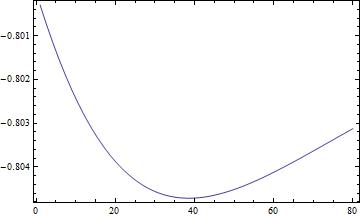}
\caption{%
The variation of $\omega(t)_{RD}$ against t, Interaction: $Q=q(3Hb\rho_{ncg}+\gamma\dot{\rho}_{ncg})$,
Parameters: $\omega_{0}=-0.8$, $\omega_{1}=-0.03$, $m=2$, $K=0.03$, $A_{0}=1$, $a_{0}=1$, $B=0.5$, $\gamma=0.1$, $b=0.3$, $\alpha=0.5$} 
\label{fig:rmemergent}
\end{figure}
Information about $\omega(t)_{RD}$ can be found in Fig. \ref{fig:rmemergent}.
\section*{\large{Intermediate scenario}}
\begin{figure}
\centering
\includegraphics[width=50 mm]{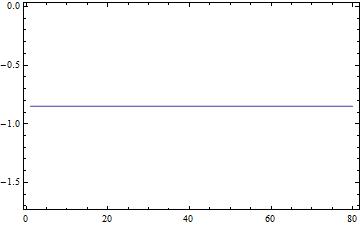}
\caption{%
The variation of $\omega(t)_{RD}$ against t,
Interaction: $Q=q(3Hb\rho_{ncg}+\gamma\dot{\rho}_{ncg})$,
Parameters: $\omega_{0}=-1$, $\omega_{1}=-0.3$, $m=0.5$, $K=1.5$, $A_{0}=1$, $\gamma=0.3$, $b=0.5$, $\alpha=0.5$} 
\label{fig:rmintermediate}
\end{figure}

\begin{figure}
\centering
\includegraphics[width=50 mm]{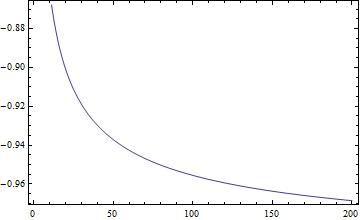}
\caption{%
The variation of $\omega_{\small{tot}}$ against t,
Interaction: $Q=q(3Hb\rho_{ncg}+\gamma\dot{\rho}_{ncg})$,
Parameters: $\omega_{0}=-1$, $\omega_{1}=-0.3$, $m=0.5$, $K=1.5$, $A_{0}=1$, $\gamma=0.3$, $b=0.5$, $\alpha=0.5$} 
\label{fig:omegaintermediate}
\end{figure}

\begin{figure}
\centering
\includegraphics[width=50 mm]{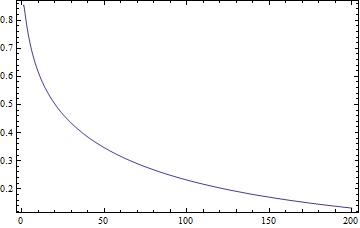}
\caption{%
The variation of $V$ against t,
Interaction: $Q=q(3Hb\rho_{ncg}+\gamma\dot{\rho}_{ncg})$,
Parameters: $\omega_{0}=-1$, $\omega_{1}=-0.3$, $m=0.5$, $K=1.5$, $A_{0}=1$, $\gamma=0.3$, $b=0.5$, $\alpha=0.5$} 
\label{fig:potintermediate}
\end{figure}

Intermediate scenario is described by $a(t)=\exp[Kt^{m}]$, with $K>0$ and $0<m<1$. Hubble parameter in this case reads as
\begin{equation}\label{eq:HubbleIntermediate}
H=Kmt^{-1+m}.
\end{equation}
$\omega(t)_{RD}$ during whole evolution of the Universe is constant
(Fig. \ref{fig:rmintermediate}) and $\omega_{\small{tot}}>-1$ corresponding to quintessence-like behavior (Fig. \ref{fig:omegaintermediate}).
Values of the parameters were chosen in order to provide $V\rightarrow0$ over time (Fig. \ref{fig:potintermediate}).
\section*{\large{Logamediate scenario}}
\begin{figure}
\centering
\includegraphics[width=50 mm]{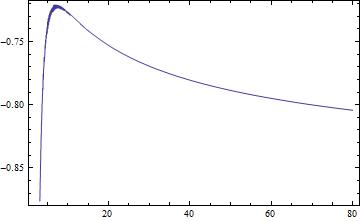}
\caption{%
The variation of $\omega_{\small{tot}}$ against t,
Interaction: $Q=q(3Hb\rho_{ncg}+\gamma\dot{\rho}_{ncg})$,
Parameters: $\omega_{0}=-0.9$, $\omega_{1}=-0.2$, $m=2$, $K=0.3$, $A_{0}=1$, $\gamma=0.1$, $b=0.4$, $\alpha=0.2$} 
\label{fig:omegalogamediate}
\end{figure}

\begin{figure}
\centering
\includegraphics[width=50 mm]{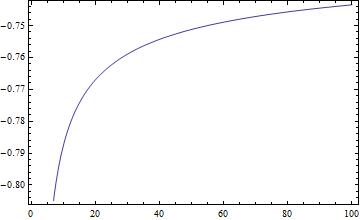}
\caption{%
The variation of $\omega_{RD}$ against t,
Interaction: $Q=q(3Hb\rho_{ncg}+\gamma\dot{\rho}_{ncg})$,
Parameters: $\omega_{0}=-0.9$, $\omega_{1}=-0.2$, $m=2$, $K=0.3$, $A_{0}=1$, $\gamma=0.1$, $b=0.4$, $\alpha=0.2$} 
\label{fig:rmlogamediate}
\end{figure}

In this section we present numerical investigation corresponding to a sign-changeable interaction between varying Generalized Chaplygin gas and Tachyonic Fluid.
Analysis of EoS parameter reveal quintessence-like behavior $\omega_{\small{tot}}>-1$ (Fig. \ref{fig:omegalogamediate}).
As in all cases, here as well, we take values for the parameters in order to
satisfy $V\rightarrow0$ condition. The behavior of time dependent EoS parameter of RD could be find in Fig. \ref{fig:rmlogamediate}.
\newpage
\section*{Discussion}
Assuming the possibility to present Darkness of The Universe as a combination of Tachyonic fluid, Generalized Chaplygin gas and leaving remaining part still untouched from mathematical point of view, we consider the following scenario. We accept that somehow, remaining darkness with time dependent EoS parameter interacts with Generalized Chaplygin gas making it varying. Interaction expressed by the fact, that EoS of Generalized Chaplygin gas becomes a function of $\omega_{RD}$ and scale factor $a(t)$. At the same time we assume that Tachyonic fluid does not feel Remaining Darkness. Having such mixture, we proposed to consider an interaction between Tachyonic fluid and  varying Generalized Chaplygin gas of the form $Q=q(3Hb\rho_{ncg}+\gamma\rho_{ncg})$, called sign-changeable. This effect mathematically possible to achieve thanks to a deceleration parameter $q$. Analysis of the problem for three different types of the scale factor was done numerically and the following was observed. In case of the scale factors describing Intermediate and Logamediate scenarios we have $\omega_{\small{tot}}>-1,$ corresponding to quintessence-like behavior. In case of Emergent scenario we have phantom-like behavior. Values for the parameters of the model were taken in order to obtain $V\rightarrow0$ over time for a potential of Tachyonic field.
\section*{Acknowledgments}
This research activity has been supported by EU fonds in the frame of the program FP7-Marie Curie Initial Training Network INDEX NO.289968.\\ \\
We would like to express our thanks to Luis P. Chimento for valuable information \cite{Chiamneto2}. 

\end{document}